# 1D Stochastic Inversion of Airborne Time-domain Electromagnetic Data with Realistic Prior and Accounting for the Forward Modeling Error

**Peng Bai** [1], **Giulio Vignoli** [1,2,*] **and Thomas Mejer Hansen** [3]

[1] Department of Civil and Environmental Engineering and Architecture (DICAAR), University of Cagliari, 09123 Cagliari, Italy; peng.bai@unica.it
[2] Geological Survey of Denmark and Greenland (GEUS), 8000 Aarhus, Denmark
[3] Department of Geoscience, Aarhus University, 8000 Aarhus, Denmark; tmeha@geo.au.dk
**\*** Correspondence: gvignoli@unica.it.

**Abstract:** Airborne electromagnetic surveys may consist of hundreds of thousands of soundings. In most cases, this makes 3D inversions unfeasible even when the subsurface is characterized by a high level of heterogeneity. Instead, approaches based on 1D forwards are routinely used because of their computational efficiency. However, it is relatively easy to fit 3D responses with 1D forward modelling and retrieve apparently well-resolved conductivity models. However, those detailed features may simply be caused by fitting the modelling error connected to the approximate forward. In addition, it is, in practice, difficult to identify this kind of artifacts as the modeling error is correlated. The present study demonstrates how to assess the modelling error introduced by the 1D approximation and how to include this additional piece of information into a probabilistic inversion. Not surprisingly, it turns out that this simple modification provides not only much better reconstructions of the targets but, maybe, more importantly, guarantees a correct estimation of the corresponding reliability.

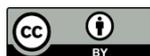

## 1. Introduction

Airborne time-domain electromagnetics (ATEM) was originally developed for mineral prospection [1–4] and is now a very consolidated methodology. Its first application probably dates back to 1948 [5], but since then, many different helicopter and fixed-wing acquisition systems have become commercially available. Amongst the most internationally known examples are: EQUATOR [6-8], Xcite [9, 10], VTEM [11, 12], SkyTEM [13], AeroTEM [14], MULTIPULSE [15], MEGATEM/GEOTEM [16, 17], TEMPEST [18, 19], and Spectrum [20].

In the last decades, technological advancements made it possible to move from the simple mineral target detection to more detailed groundwater mapping [21, 22] and geological modeling applications [23, 24].

Together with the enhanced capabilities of the instrumentation, the strategies to deal with the collected data have also been largely improved. For example, similarly to what happens in seismics (e.g. [25]), the stacking of the recorded transient curves—by means of moving windows whose widths depend on the time-gate [26–28]—can be used to preserve high lateral resolution at shallow, while increasing the reliability of the data at depth; this is, indeed, compatible, at the same time: (i) with the higher signal-to-noise at the early-gates, and also (ii) with the larger spatial footprint at the late-gates.

Moreover, novel strategies to effectively incorporate prior information into the translation of the observed data into conductivity models have led to the development of, for example, spatially constrained inversion schemes [29-32] (again, similarly to approaches utilized for other, very different, kinds of data [33–37]). This capability of enforcing spatial coherence allowed the reconstruction of (pseudo-)3D conductivity distributions even by means of simple 1D forward modeling [38, 39].

Clearly, approaches based on 1D forward modeling are particularly convenient for their computational performances [40]. Recent strategies exploit the neural networks potential, allowing almost real-time inversions with no significant quality reductions [41]. In addition, in the attempt of retrieving, not merely the conductivity distribution, but, rather, immediately useful pieces of information about the targets and, at the same time, in the effort of supplying reliable assessment of the associated uncertainty, probabilistic petrophysical inversions are becoming more and more common [42].

Unfortunately, as mentioned, all these approaches rely on 1D approximation of the physical phenomena. Very often, the error introduced by using this rough description of the reality is neglected, and when, in the most conservative workflows, it is not, the modeling error is accounted for by arbitrarily increasing the data fitting threshold [43, 44]. Of course, disregarding this further noise component during 1D inversions could potentially be extremely dangerous as fitting the 3D observations with 1D forward modeling might produce reconstructions appearing very detailed, and, seemingly, characterized by high resolution, but that, in reality, are affected by correlated noise artifacts [45]. The necessity of tackling three-dimensionality issues in effective and pragmatic ways is even more crucial nowadays as large datasets are (and are going to be) collected for (ultra-)shallow mapping [46] where even sedimentary environments can be characterized by extremely high heterogeneity; in such a case, 2D/3D artifacts, due to improper modeling error assessment, would pervade the conductivity (and/or petrophysical) reconstructions without being detrimental in terms of the data fitting levels [47]. Consequently, the results could nicely fit (in a 1D sense) the (3D) observed data and potentially seem very accurate, but actually be permeated by pitfalls.

In principle, the most straightforward way to deal with these dimensionality problems would consist of using fully 3D forward modeling algorithms. Indeed, much research is currently devoted to practical 3D inversion approaches [1, 39, 48–54]. However, 3D approaches are still too computationally demanding for being routinely used [55], especially within probabilistic frameworks [56].

In the present research, we show that the 1D modeling error can be significant and lead to erroneous/misleading inversion results. In addition, we discuss how such modeling error can be quantified based on the available prior information about the system under investigation. Once quantified, this resulting noise model could be incorporated, in principle, in both deterministic and probabilistic inversions to correctly account for the 1D approximation. Finally, we demonstrate how using such a correlated noise model positively impacts even the probabilistic inversion of ATEM data.

We test the proposed approach on synthetic datasets characterized by increasing complexity levels. Through them, we demonstrate the effectiveness of the proposed approach: (i) in reconstructing features that are indeed in the 3D data, and (ii) in providing the crucial, associated uncertainties. For each of the considered Tests, the results obtained via: (i) a "standard" deterministic 1D Occam's inversion, (ii) a stochastic inversion with realistic prior, and (iii) the same stochastic inversion, but now taking into account the modeling error consistent with the used prior are compared.

## 2. Methodology

The calculation of the response of a physical system is always characterized by some level of modeling error. For example, even in the case of very sophisticated 3D forward modeling tools used to calculate the $dB/dt$ responses of an electrical conductivity distribution caused by the excitation induced by a "perfectly" described ATEM system (is it really possible to characterize a priori an acquisition system? [8, 13, 44]), the parameterization used and the size of the discretized domain might affect the retrieved response. In this respect, every time we use a 1D forward modeling approximation [43] for the inversion of ATEM data (inherently 3D), we introduce some errors that need to be handled; neglecting this source of additional uncertainty would inevitably lead to artifacts, paving the road to successive geological misinterpretations. In the best scenario, the magnitude of the modeling error will be negligible compared to uncertainty in the data, and can be ignored [57]. More often, modeling errors will be significant, and must be taken into account.

Here, we briefly recall a formal framework in which the modeling error is described through a multivariate normal correlated probability distribution, which can be naturally used when the measurement errors are also Gaussian [57]. In probabilistic formulations of inverse problems, the goal is to retrieve the posterior probability density function $p(\mathbf{m}|\mathbf{d})$ measuring the probability of having the model $\mathbf{m}$ compatible with the measurements $\mathbf{d}$. In accordance with Bayes' theorem, $p(\mathbf{m}|\mathbf{d})$ is proportional to the product between the prior probability density function of the model parameters $p(\mathbf{m})$, and the conditional probability density function $p(\mathbf{d}|\mathbf{m})$. Hence, $p(\mathbf{m}|\mathbf{d}) \propto p(\mathbf{d}|\mathbf{m})p(\mathbf{m})$, with $p(\mathbf{d}|\mathbf{m})$ connecting the measured data and the model parameters, and, in the specific case of a Gaussian noise distribution, can be written as

$$p(\mathbf{d}|\mathbf{m}) = k_d \exp\left(-\frac{1}{2}\big(\mathbf{d} - F(\mathbf{m})\big)^T \mathbf{W}_d^T \mathbf{W}_d \big(\mathbf{d} - F(\mathbf{m})\big)\right), \quad (1)$$

where: (i) $k_d$ is just a normalization factor, (ii) $F$ is the forward modeling operator used during the inversion, and (iii) $\mathbf{W}_d$ is related to the data covariance $\mathbf{C}$, and, often, by assuming mutually independent data, can be considered equal to $\mathbf{W}_d = \text{diag}(\mathbf{\sigma}_d)^{-1} = \mathbf{C}^{-1/2}$ where the $i$-th component of the vector $\mathbf{\sigma}_d$ is the standard deviation of the $i$-th data component (in the specific case of the ATEM data, $[\mathbf{\sigma}_d]_i$ is the standard deviation of the $dB/dt$ value associated with the $i$-th time-gate).

If the model parameters are also assumed to follow a Gaussian distribution, then the prior information about the solution can be formalized as follows: $p(\mathbf{m}) = k_m \exp\left(-\frac{1}{2}(\mathbf{m} - \mathbf{m}_0)^T \mathbf{W}_m^T \mathbf{W}_m (\mathbf{m} - \mathbf{m}_0)\right)$, in which: (i) $k_m$ is another normalization factor and (ii) the Gaussian is centered on the reference model $\mathbf{m}_0$. In this specific case of Gaussian distributions assumed both for the data noise and the model parameters, we obtain that the maximizer of the probability $p(\mathbf{m}|\mathbf{d})$ is also the minimizer of the regularized inversion objective functional $\|\mathbf{W}_d(\mathbf{d} - F(\mathbf{m}))\|_{L_2}^2 + \|\mathbf{W}_m(\mathbf{m} - \mathbf{m}_0)\|_{L_2}^2$. In fact, this is a very well-known result (e.g., [36, 58-59]), to some extent, reconciling probabilistic and deterministic approaches; in particular, if $\mathbf{C}_m^{-1}$ is taken equal to $\lambda^2 \mathbf{L}^T \mathbf{L}$—with $\lambda$ being the Tikhonov parameter controlling the relative importance of the regularization term with respect of the data misfit, and $\mathbf{L}$ a discrete approximation of the spatial derivative—then, the minimization of the objective functional coincides with the standard Occam's inversion [60].

However, the approach discussed in the present research loosens several of these ansätze, and, in the following:

i. we do not restrict ourselves to the Gaussian assumption for the model parameters distribution $p(\mathbf{m})$ as we are going to consider quite general prior distributions defined through the realizations of those distributions and that will be generated via a geologically informed procedure;
ii. the $\mathbf{W}_d = \mathbf{C}^{-1/2}$ will not consist uniquely of the component attributable to the noise in the observations, but it will also include a term incorporating the modeling error. In particular, the modeling error will be assumed to be consistent with a Gaussian probability density $\mathcal{N}(\mathbf{d}_\Delta, \mathbf{C}_\Delta)$ defined by the mean $\mathbf{d}_\Delta$ and the covariance $\mathbf{C}_\Delta$. Hence, the $p(\mathbf{d}|\mathbf{m})$ in Equation 1 will have now the following expression [47]

$$p(\mathbf{d}|\mathbf{m}) = k_d \exp\left(-\frac{1}{2}(\mathbf{d} + \mathbf{d}_\Delta - F(\mathbf{m}))^T \mathbf{W}_\Delta^T \mathbf{W}_\Delta (\mathbf{d} + \mathbf{d}_\Delta - F(\mathbf{m}))\right), \quad (2)$$

with $\mathbf{W}_\Delta = (\mathbf{C} + \mathbf{C}_\Delta)^{-1/2}$.

By construction, as it will be detailed in what follows, these new terms $\mathbf{d}_\Delta$ and $\mathbf{C}_\Delta$ will also depend on the prior geological knowledge available about the investigated area.

It is important to stress that despite us demonstrating the effects of taking into account the modeling error within a probabilistic framework, the $\mathbf{W}_\Delta = (\mathbf{C} + \mathbf{C}_\Delta)^{-1/2}$ can actually be, in a very immediate way, incorporated into a deterministic framework as well. However, the evident advantage of using a stochastic approach is that we can naturally incorporate complex prior information (rather than enforcing simple—e.g., smooth or sharp—constraints) and those pieces of complex information need to be available in any case since they are used for the assessment of the modeling error.

Moreover, given the arbitrariness we can benefit from by defining the prior distribution directly via its samples, in general, we could potentially have the maximum flexibility and even use very powerful strategies as, for example, those based on Multiple-point statistics (MPS) approaches [61]; in those cases, the prior geological information can be formalized by means of the so-called Training Image (TI) that is, basically, representing the conceptual geological model of the expected target subsurface; MPS algorithms can generate samples of the prior that are statistically stationary with respect to the original TI and that can be used as detailed in the following Subsection. In the present research, however, we use other geostatistical strategies to populate the prior (and consistently estimate the associated modeling error). They will be described in the following Section 3 "Results".

*2.1. Estimation of Gaussian Correlated Modeling Errors*

Here, we follow the strategy detailed in [47] to (i) simulate and (ii) quantify modeling errors caused by using a 1D forward as opposed to a full 3D forward for simulating ATEM data.

Firstly, a sample of the underling probability distribution representing the modeling errors is generated. This is done by generating a relatively large sample of $N_\Delta$ realizations of the prior distribution $p(\mathbf{m})$ as $\mathbf{M} = [\mathbf{m}'_1, \mathbf{m}'_2, \ldots, \mathbf{m}'_{N_\Delta}]$. The forward response is then calculated by using the approximate 1D forward model, $F_{app}$, and the (assumed) exact 3D forward model, $F_{ex}$. This provides a set of 'approximate' and 'exact' data in the form of $\mathbf{D}_{app} = [F_{app}(\mathbf{m}'_1), F_{app}(\mathbf{m}'_2), \ldots, F_{app}(\mathbf{m}'_{N_\Delta})]$ and $\mathbf{D}_{ex} = [F_{ex}(\mathbf{m}'_1), F_{ex}(\mathbf{m}'_2), \ldots, F_{ex}(\mathbf{m}'_{N_\Delta})]$. Hence, the difference between the approximate and the exact forward models represents a realization, $[\mathbf{D}_{diff}]_i = [\mathbf{D}_{ex} - \mathbf{D}_{app}]_i$, of the modeling error associated to the specific i-th sample of the prior. As will also be discussed in the following, to feed the 1D forward $F_{app}$, unidimensional conductivity models have been extracted from the original 3D realizations of the prior in the locations just below the acquisition system positions.

Assuming that the modeling error can be characterized by a multivariate Gaussian distribution $\mathcal{N}(\mathbf{d}_\Delta, \mathbf{C}_\Delta)$, the mean and covariance can be trivially computed from the samples $\mathbf{D}_{diff}$.

Finally, the assessment of the 1D modeling error can be plugged into Equation 2.

The number of realizations needed, $N_\Delta$, and the validity of the Gaussian assumption on the modeling errors will be addressed below.

In the present research, as the best approximation $F_{ex}$, it has been considered an implementation of the forward modeling discussed in detail in [62]. The simulations have been preceded by a validation phase, in which the 3D forward modeling results have been compared against known solutions assumed to be exact. In our specific case, we performed preliminary tests of the 3D forward against semi-analytical solutions for unidimensional conductivity distributions; the chosen 3D simulation settings have led to mismatches of a few percentage points (generally around 5%). Consequently, as it will be shown later in the paper, the modeling error inherited from the 3D modeling has been assumed negligible with respect to the other noise sources and it has not been further considered in our analysis.

*2.2. Inversion Strategies*

Concerning the deterministic inversion results discussed in the following, as mentioned, we use an Occam's inversion scheme [63] in which each individual $dB/dt$ sounding is inverted independently from the adjacent ones and, therefore, the roughness operator in the regularization terms acts only vertically. To be fair, it is true that this specific kind of prior information, formalized by the stabilizer, is not in accordance with the investigated models (as it will be clear in the descriptions of the tests, characterized by abrupt conductivity changes). Nevertheless, the deterministic algorithm retrieves (smooth) models whose 1D responses are in almost perfect agreement with the inverted 3D $dB/dt$ data. The level of data fitting used for the inversion of the noise-free data is, for all tests, 0.01%.

The stochastic inversion consists of an independent extended Metropolis algorithm, which is analogous to the standard extended Metropolis algorithm [64], but with an infinitely long step-length; this choice implies that every new model proposal from the prior is a completely independent realization. Accordingly, its implementation consists of a slight modification of the SIPPI toolbox [65, 66], making the proposed samples independently drawn from the prior ensemble. The $10^5$ realizations of the prior are generated in advance since a subset of them needs to be used for the modeling error assessment; the appropriateness of this choice, and the convergency properties will be discussed in the section "Discussion".

**3. Results**

In this section, we perform two synthetic tests of increasing complexity to investigate the effects of including in the inversion process: (i) the proper prior information, and (ii) an estimation of the modeling error. In both cases, we compare our results against a solution provided by a more standard 1D deterministic inversion.

*3.1. Test 1: 3D Conductivity Distribution with Homogeneous Layers*

The first test (Test 1—Figure 1) consists of a conductivity distribution mimicking possible glacial geological settings typical, for example, of Denmark [24, 67-68] and characterized by an intricated network of paleovalleys.

Figure 2 compares, for the sounding locations considered in Figure 1, the 3D responses calculated for the entire 3D model against the corresponding 1D measurements that would be obtained by considering exclusively the 1D portion of the original distribution just below the acquisition position. The acquisition specifications are those of a typical VTEM system [11, 12] (e.g., each sounding consists of 54 measurements). Not surprisingly, the 1D responses (Figure 2b) are characterized by abrupt lateral changes associated with the lateral variations of the conductivity model (Figure 2a), whereas, in accordance with the physics, the 3D $dB/dt$ data are much smoother (Figure 2c). Here, we treat the 3D calculated responses as the 'observed data' and invert them by means of both deterministic and probabilistic inversion methods.

In all cases we assume negligible measurement errors during the inversion, as the main purpose is to focus on the modeling error effects. In this respect, it is important to highlight the level that the modeling error can reach: the difference in the 1D and 3D responses (blue line in Figure 2b) can be as big as ~20%, and, very seldom, smaller than several percentage points. Therefore, the size of this mismatch should make accounting for the modeling error unavoidable; on the contrary, as mentioned before, the common practice is to tackle it by discretionally increase the measurement noise. By properly including the modeling error in the inversion, the assumed measurement error could be potentially reduced.

### 3.1.1. Deterministic Occam's Inversion

Figure 3 shows the result of 1D deterministic inversion—implementing an Occam's regularization strategy. Clearly, the deterministic result fits the observations extremely well (blue line in Figure 3b) and is capable to retrieve the major features of the true model whereas, in some cases, infers deceptive discontinuous reconstructions of the true interfaces (e.g., ~1600 to ~2000 m). It is worth mentioning that the data sensitivity to the model parameters drops below the first reconstructed conductive interface; hence, eventually, all the conductivity variation below that retrieved deep interface would not be considered reliable by skilled interpreters. In this respect, the deterministic result cannot be considered, at least from a practical point of view, much different from the probabilistic result obtained without accounting for the modeling error (and discussed in depth in the following Subsection). Nevertheless, it is true that the retrieved (laterally discontinuous) features (caused by fitting the coherent modeling error—as it will be clear later) might be challenging to be correctly deciphered and might lead to erroneous conclusions.

### 3.1.2. Stochastic Inversion without Modeling Error Assessment

The stochastic approach allows incorporation of (in principle) arbitrarily complex prior information, as long as realizations from the prior can be generated. In the present research, we generated independent realizations of $p(\mathbf{m})$ representing buried valley structures (similar to those in Figure 1) by means of a Fast Fourier Transform Moving Average (FFT-MA) strategy [69] providing unconditional realizations of a Gaussian random field of the interfaces' locations; in particular, the corresponding mean values are chosen to be uniformly distributed: (i) between 20 and 30 m for the shallowest interface, and (ii) between 65 and 85 m for the deepest, whereas the associated covariance models are characterized by a standard deviations and ranges, respectively, of: (i) 5 m and 100 m, for the first interface, and (ii) 80 m and 500 m for the second.

More details about possible ways of constructing the samples can be found in [65, 66] including, for example, some details about the above-mentioned MPS strategies.

Whilst the geometry is varying, the conductivity values of each layer are kept constant realization-by-realization. For clarity, two examples of prior distribution realizations can be seen in Figure 4. It is evident that the parameters defining the realizations of the prior have been selected to be in agreement with our expectations about the geological structures we are dealing with.

It is worth highlighting that, since the inversion is one-dimensional, the actual (1D) models used to feed the inversion algorithm are the individual columns (and the corresponding 1D response) of each 3D realization (e.g., in Figure 4). Therefore, each single 1D prior model can be described by five parameters (the depths to the interfaces and the three conductivities).

Undoubtedly, the stochastic inversion, with such an informative prior, is quite facilitated and is basically reduced to the inference of the locations of the interfaces. As a matter of fact, as Figure 5 shows, by using the Metropolis algorithm [64, 70], we can reconstruct quite satisfactorily the first interface at about 25-30 m depth, whereas we generally underestimate the depth of the second layer (except for the last ~1500 m on the right). However, what is most disturbing is that, even in this simple case, the misleading features reconstructed by the stochastic inversion (here, performed without accounting for the modeling error) appear to be almost certain.

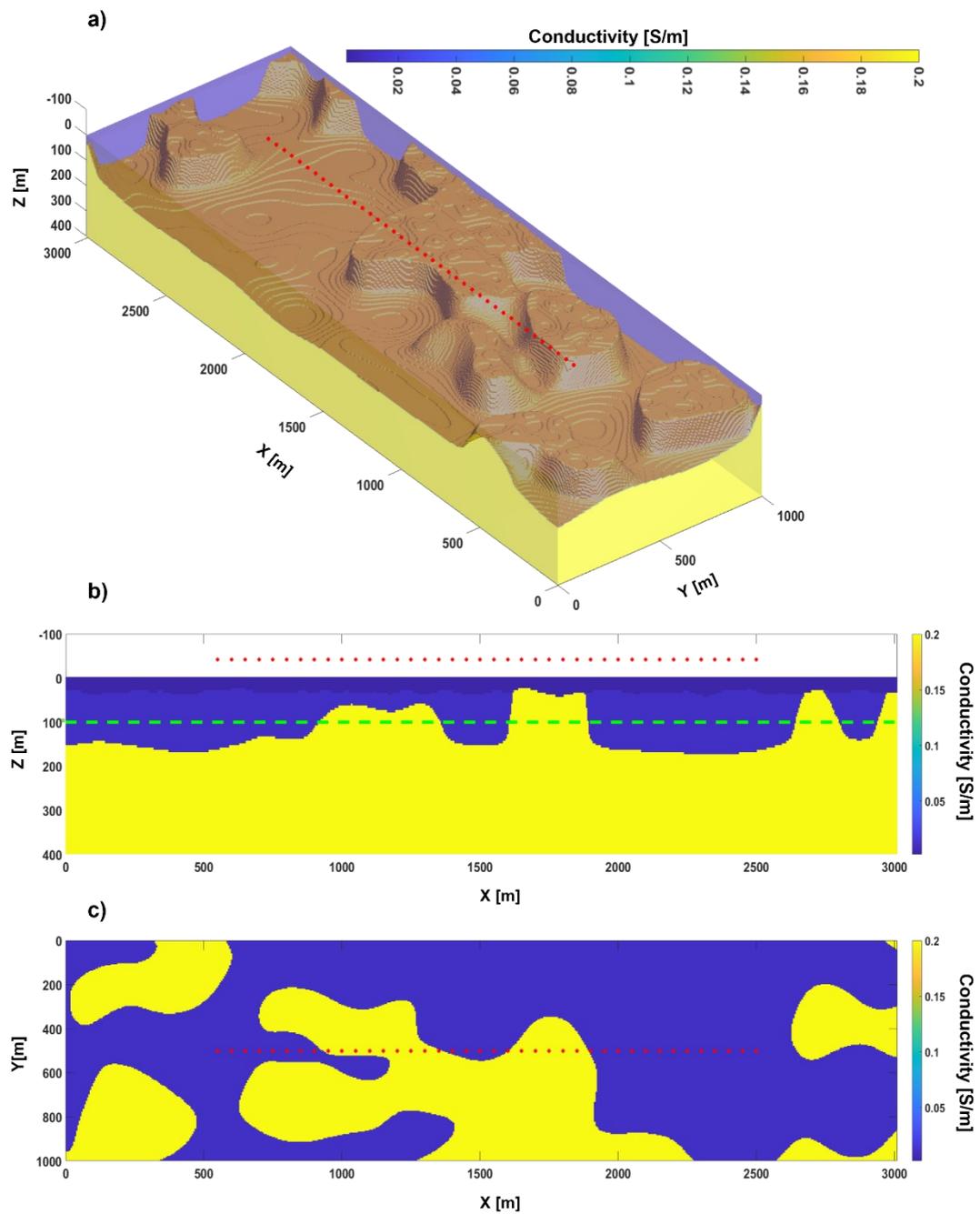

**Figure 1.** Conductivity distribution for Test 1: (**a**) 3D view of the model consisting of a sequence of three homogeneous layers with varying thicknesses; (**b**) Vertical section of the model in panel (**a**), along the survey line highlighted by the red dots indicating the locations of the ATEM soundings; (**c**) Plain view at 100 m depth (dash green line in panel (**b**)).

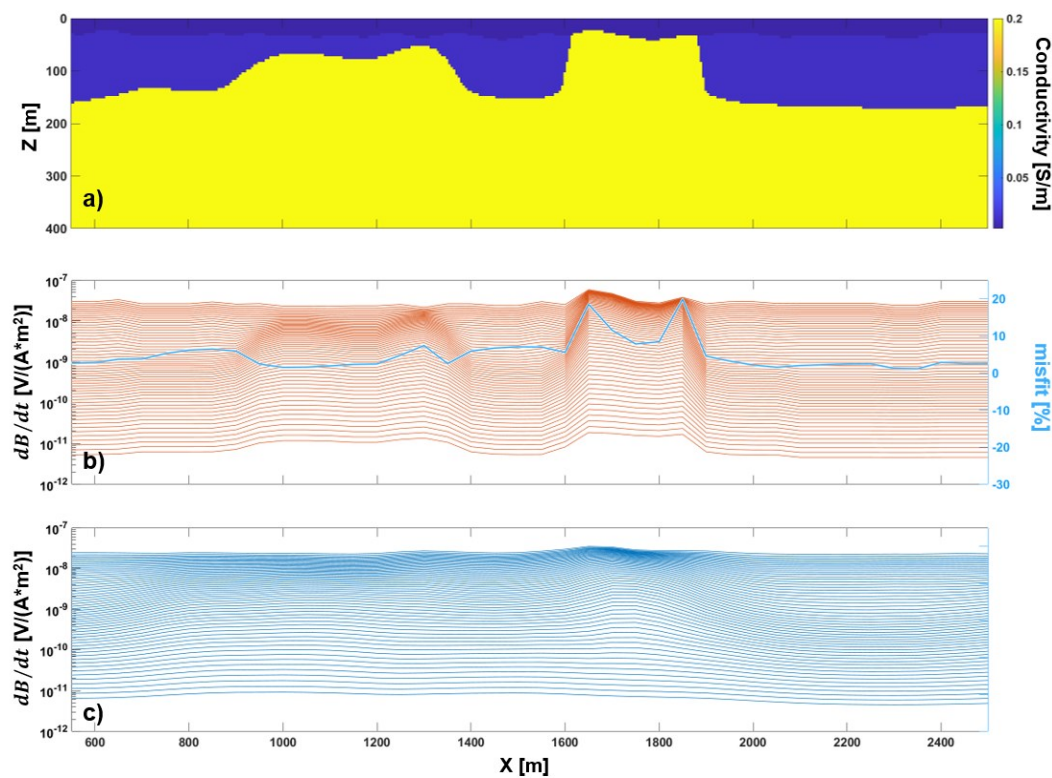

**Figure 2.** Comparison of the 3D and 1D responses for the conductivity model of Test 1: (**a**) Vertical section of the 3D conductivity model in Figure 1a (it is a portion of the section in Figure 1b); (**b**) 1D responses calculated for the 1D portions of the original model in Figure 1a taken at the locations of the ATEM soundings (red dots in Figure 1b) – the blue line represents, sounding-by-sounding, the relative misfit between the 1D and 3D responses (the corresponding axis is on the right, in blue); (**c**) 3D responses measured at the same location in panel (**b**), but, here, calculated for the entire 3D conductivity model (Figure 1a).

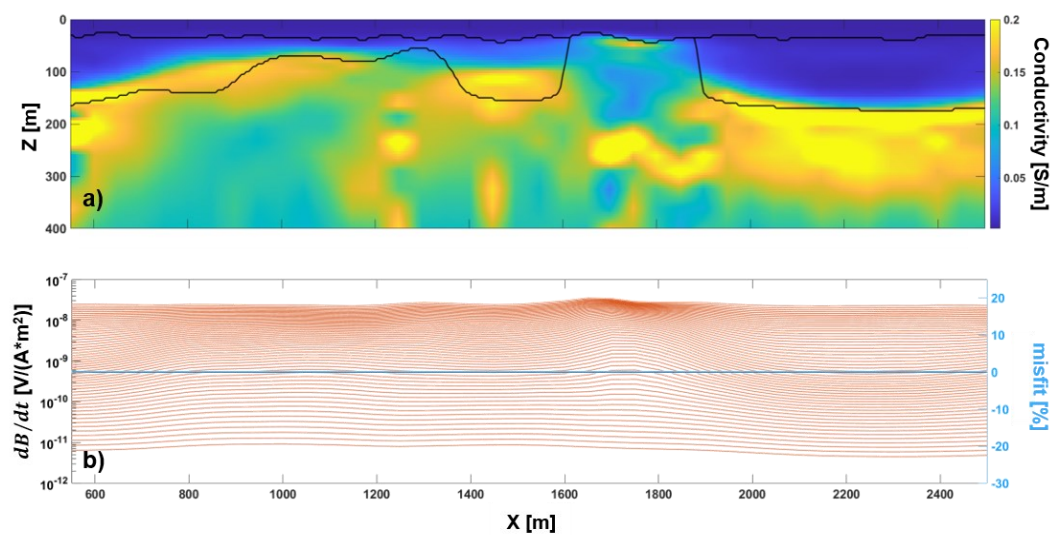

**Figure 3.** 1D deterministic inversion of the 3D data (Figure 2c) associated to the conductivity model of Test 1 (Figure 1a): (**a**) The solution of the 1D deterministic inversion – the black lines show the interfaces of the original conductivity model to be reconstructed; (**b**) The 1D responses resulting from the conductivity model in panel (**a**) – the blue line represents sounding-by-sounding, the relative misfit between the 3D and 1D responses (the corresponding axis is on the right in blue).

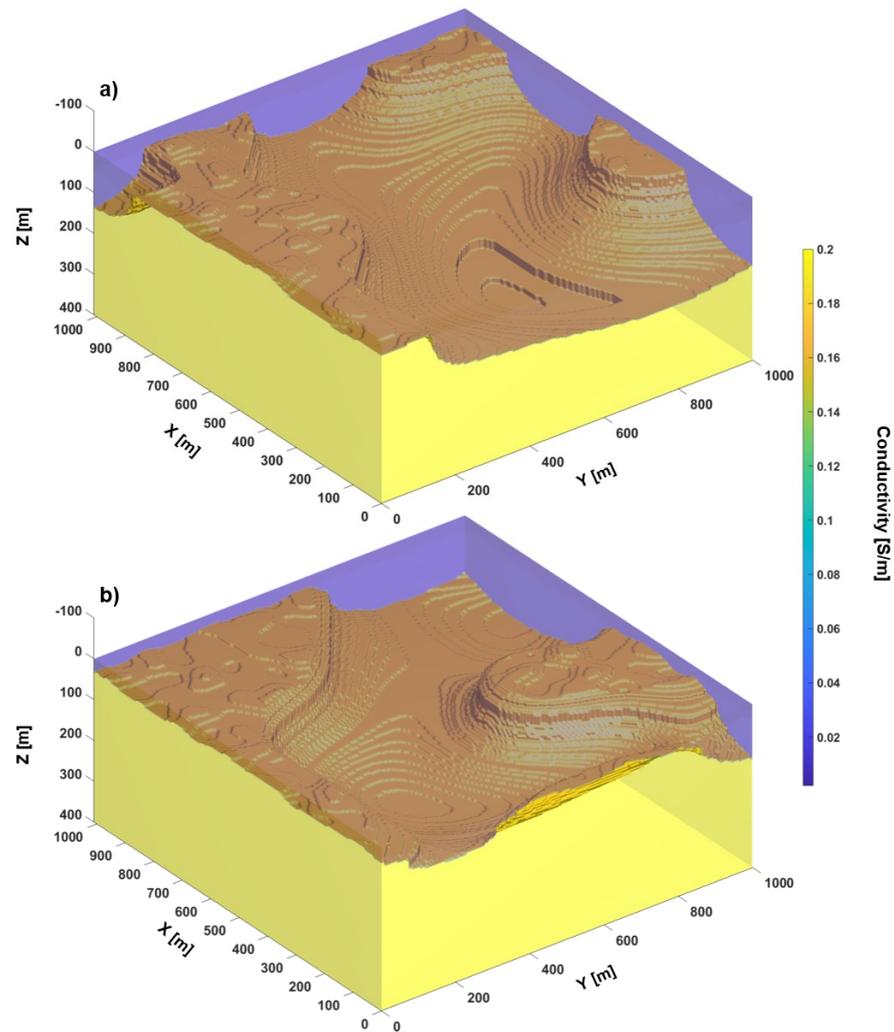

**Figure 4.** (**a**) and (**b**) show two examples of possible, distinct, realizations of the prior distribution used for the stochastic inversion of Test 1's dataset.

3.1.3. Stochastic Inversion Incorporating the 1D Modeling Error

If instead, we use a subset of the 3D prior samples (Figure 4) to calculate their actual 3D responses and compare them with the data calculated, this time, by means of the 1D forward modeling applied to the 1D conductivity vertical profile in the center of each selected prior realization, we can estimate the appropriate mean $\mathbf{d}_\Delta$ and covariance $\mathbf{C}_\Delta$, and incorporate them into the 1D stochastic inversion scheme. The results of the application of this new scheme to Test 1 is shown in Figure 6. Now, the result almost perfectly matches the true model. When the mean depth of the interfaces does not fit the true conductivity change—for example, near the steep lateral variations around 1000 m and 1400 m—the correct location of the boundaries still lies within the uncertainty bands.

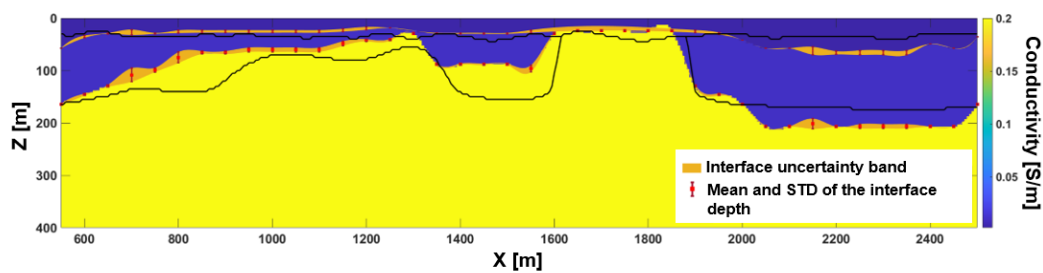

**Figure 5.** 1D stochastic inversion of the 3D data (Figure 2c) associated to the conductivity model of Test 1 (Figure 2a shows one vertical section of that 3D conductivity model). In this case, realistic prior is used (Figure 4), but no modeling error has been taken into account. The orange bands represent the reconstruction uncertainty defined by the mean and standard deviation values (red points and vertical bars) deduced by the retrieved realizations of the posterior $p(\mathbf{m}|\mathbf{d})$. As for Figure 3a, the black lines show the locations of the interfaces to be reconstructed.

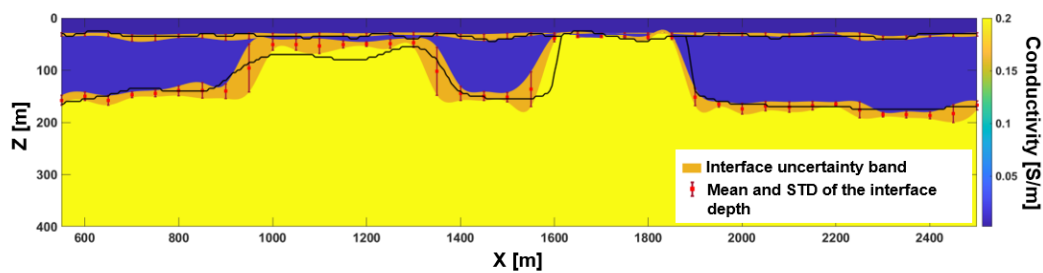

**Figure 6.** 1D stochastic inversion of the 3D data (Figure 2c) associated with the conductivity model of Test 1 (Figure 1a). In this case, realistic prior is used (Figure 4) together with the modeling error assessment. As in Figure 5, the orange bands represent the uncertainty (the associated mean and standard deviations are plotted as vertical red bars) calculated by the retrieved realizations of the posterior $p(\mathbf{m}|\mathbf{d})$. The black lines show the locations of the interfaces to be reconstructed (Figure 2a).

Since we are dealing with 3D conductivity models, it is probably more appropriate to visualize the consequences of the different inversion schemes over several flight lines (Figure 7) across the Test 1 model. If we examine Figures 7b,c, the same conclusions drawn for an individual vertical section are clearly valid also for the other acquisition lines (and, consistently, also for the lateral intra-line resolution—kindly, compare Figures 7e,g).

A direct assessment of the performances of the different inversion schemes in terms of data fitting can be performed by looking at Figure 8b; in general, the relative mismatch between each of the channels of the observed 3D responses and those calculated (via a 3D forward) from the 3D conductivity distribution obtained via the stochastic inversion (with modeling error appraisal) lays within $\mp 10\%$. It is worth noticing that the relatively poor data fitting on the right side of Figure 8b is caused by the higher level of heterogeneity of that end of the model, beyond the surveyed area. In fact, Figure 7d shows the rapidly varying morphology for $X > 2500$ m in correspondence of the section considered in Figure 8 ($Y \sim 500$ m); that conductivity variation is not in accordance with the inevitable laterally homogeneous extension of the reconstructed solution (Figure 7g). This mismatch affects, not surprisingly, mainly the late-gate measurements. The same does not happen on the other end of the section, where the data fitting is particularly good (Figure 8b); in this case, the reason is that, differently from before, the true model continues largely unchanged towards low $X$ values, at least for $Y \sim 500$ m (Figure 7d). On the other hand, the 3D responses generated by the conductivity volume retrieved by the 1D deterministic inversion (Figure 8a) demonstrates, once more, that fitting the data with a 1D forward (as in Figure 3b) does not necessarily guarantee that the corresponding 3D calculated data are in agreement with the (3D) observations; indeed, in Figure 8a, the relative misfit between the 3D calculated responses and the measurements ranges approximately between ∓30%.

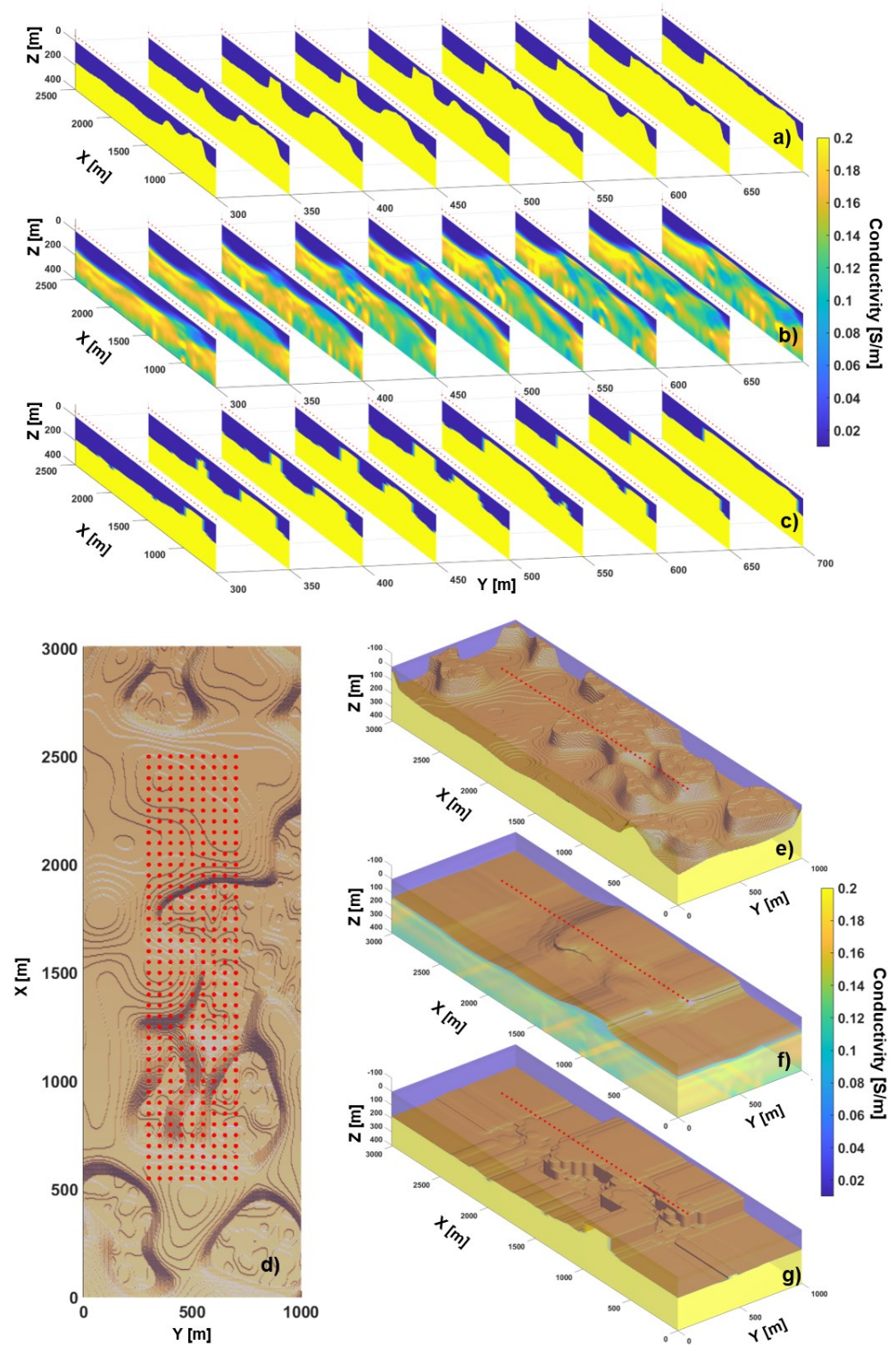

**Figure 7.** The inversion results of the 3D data from (**a**) Test 1's conductivity model (Figure 1) and obtained with: (**b**) Occam's deterministic strategy, and (**c**) The stochastic strategy incorporating the modeling error. The locations of the inverted soundings are shown as red dots in (**d**) on top of the plain view of the topography of the second interface of the true model – shown in (**e**). In (**f**), the 3D view of the results from (**b**) is plotted, whereas, in (**g**), it is possible to see a similar 3D view but now based on the stochastic results in (**c**). For clarity, in panels (**e** – **g**), only the sounding locations of the central flight line are shown (as red dots).

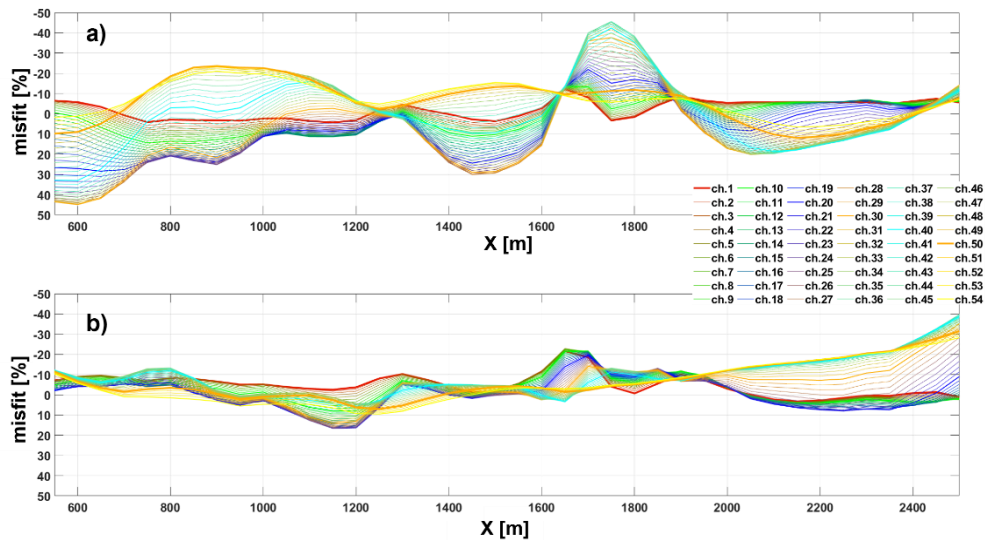

**Figure 8.** Comparison of the percentage misfit between the observed 3D responses and those calculated from: (**a**) The 3D conductivity model retrieved by the 1D deterministic approach (Figures 7b and 7f); (**b**) The 3D conductivity model obtained via the stochastic inversion accounting for the modeling error (Figures 7c and 7g). Each of the 54 plotted time-gates (channels) is depicted with a different color.

*3.2. Test 2: 3D Conductivity Distribution with Heterogeneous Layers*

In Test 2, we apply the same strategy to a more elaborated 3D conductivity model. Test 2's model consists of three layers with similar geometries as in Test 1, but, now, characterized by heterogeneous conductivity values (Figure 9). For sake of completeness, analogously to what has been done for Test 1, also for Test 2, we show the 3D responses (Figure 10c) along the central profile (Figure 10a) of the 3D conductivity model (Figure 9a). In Figure 10b, we display the 1D response as they would result by considering each column of the true conductivity model as independent (for comparison, kindly, see Figure 2b concerning Test 1). Clearly, for Test 2, the importance of taking into account the modeling error is even more evident: the error we introduce when interpreting the 3D model response in terms of 1D data is never below 4% (blue line in Figure 10b).

Figure 11a presents the result of Occam's inversion of the 3D data (Figure 10c). Again, from Figure 11b, it is clear that it is not difficult to perfectly fit the 3D data with 1D responses, even with a model barely capable to get the very major features of the conductivity model to be inferred.

Also for this more complex test, a 1D stochastic inversion can be performed by following the previously discussed Metropolis approach, making use of precalculated samples. The $10^5$ realizations of the prior distribution are very similar to those shown in Figure 4 for Test 1; the only difference is that, consistently with the new model to be reconstructed, also for the prior realizations, the conductivity of each layer is allowed to vary.

Figure 12 shows the mean map of the posterior distribution obtained without taking into consideration the modeling error; as for Figure 5 (about Test 1), also for Test 2, the retrieved mean model (Figure 12a) does not capture the lateral variation of the top of the deepest layer and smooths the large majority of the incisions out. Perhaps worse than that is the fact that, as shown by the standard deviations of the conductivity of each layer (plotted in Figure 12b), the leveled-out reconstruction is given for (almost) undeniable.

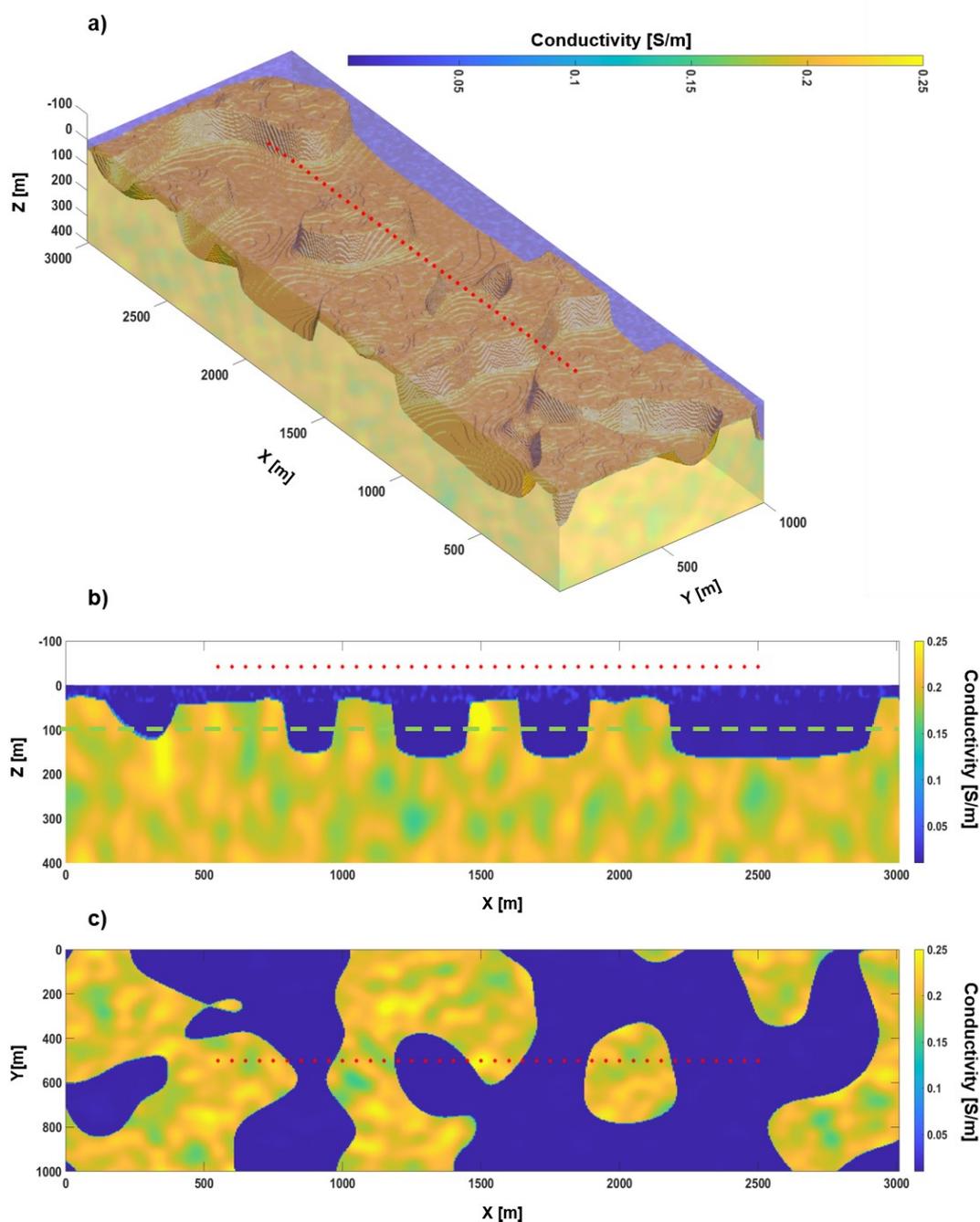

**Figure 9.** Conductivity distribution for Test 2: (**a**) 3D view of the model consisting of a sequence of three heterogeneous layers with varying thicknesses; (**b**) Vertical section of the model in panel (**a**), along the survey line highlighted by the red dots indicating the locations of the ATEM soundings; (**c**) Plain view at 100 m depth (dash green line in panel (**b**)).

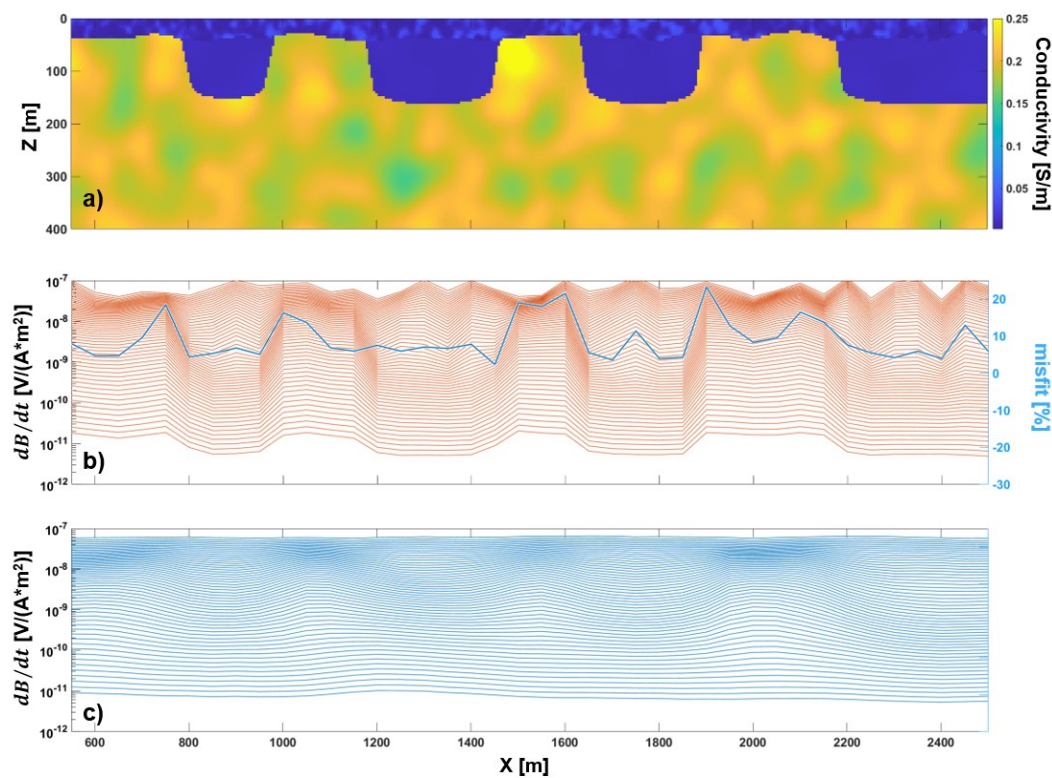

**Figure 10.** Comparison of the 3D and 1D responses for the conductivity model of Test 2: (**a**) Vertical section of the 3D conductivity model in Figure 9a (it is a portion of the section in Figure 9b); (**b**) 1D responses calculated for the 1D portions of the original model in Figure 9a at the locations of the ATEM soundings (red dots in Figure 9b) – the blue line represents sounding-by-sounding, the relative misfit between the 1D and 3D responses (the corresponding axis is on the right in blue); (**c**) 3D responses measured at the same location in panel (**b**), but, here, calculated for the entire 3D conductivity model (Figure 9a).

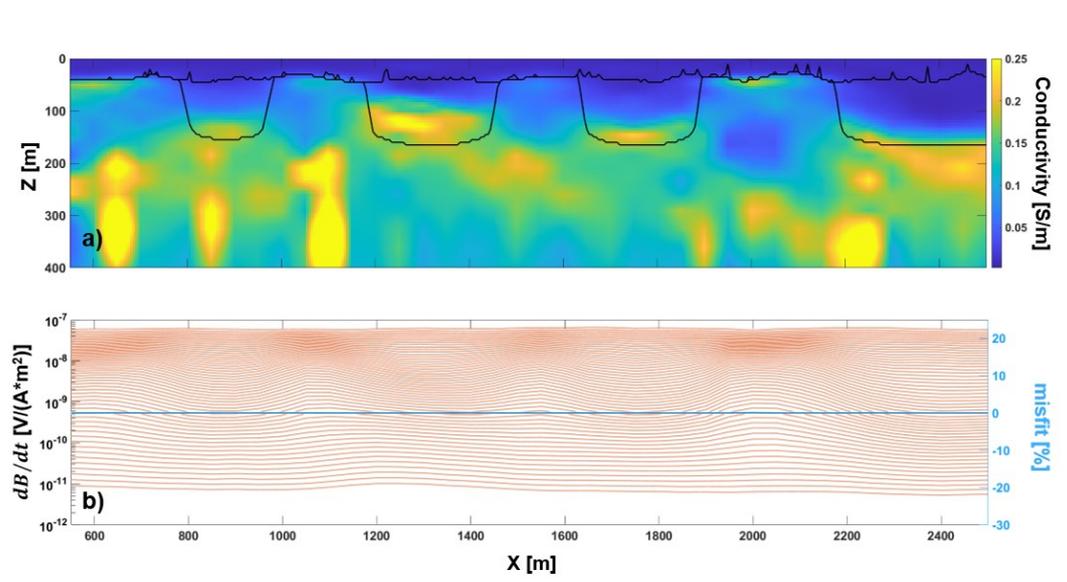

**Figure 11.** 1D deterministic inversion of the 3D data (Figure 10c) associated to the conductivity model of Test 2 (Figure 9a): (**a**) The solution of the 1D deterministic inversion – the black lines show the interfaces of the original conductivity model to be reconstructed; (**b**) The 1D responses resulting from the conductivity model in panel (**a**) – the blue line represents, sounding-by-sounding, the relative misfit between the 3D and 1D responses (the corresponding axis is on the right in blue).

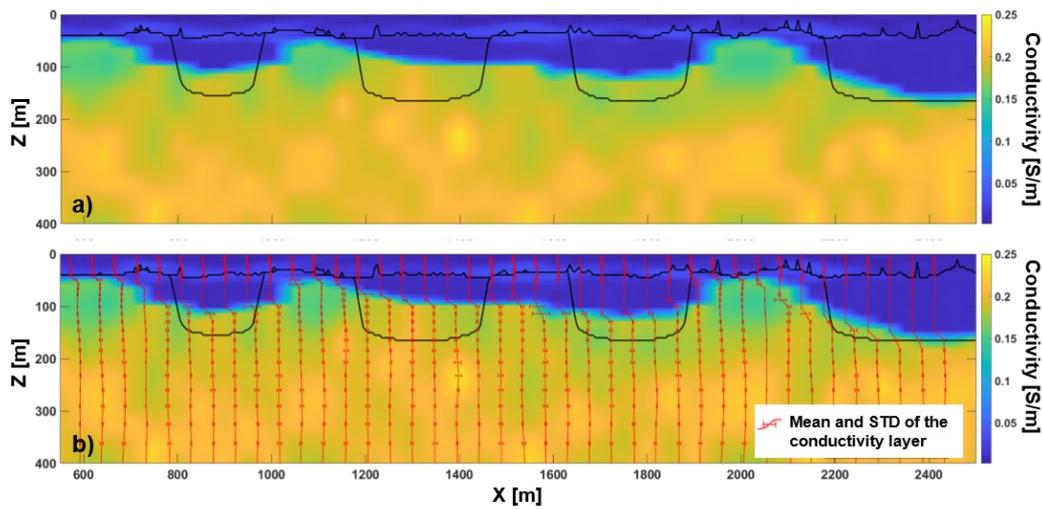

**Figure 12.** (**a**) Mean map of the 1D stochastic inversion of the 3D data (Figure 10c) associated to the conductivity model of Test 2 (Figure 10a shows one vertical section of that 3D conductivity model). In this case, realistic prior is used, but no modeling error has been taken into account. The black lines show the locations of the interfaces to be reconstructed. (**b**) The same reconstructed conductivity distribution as in panel (**a**), but now with the mean and standard deviation vertical profile superimposed for each vertical conductivity profile.

On the contrary, if we take into account the modeling error, the mean map derived from the 1D stochastic inversion provides a quite satisfactory reconstruction of the true model with all its complex morphology (Figure 13a), and when the inferred mean conductivity values do not reflect the true distribution, they are associated with high levels of uncertainty (for example at depth ~100 m and $X$~1300 m in Figure 13b).

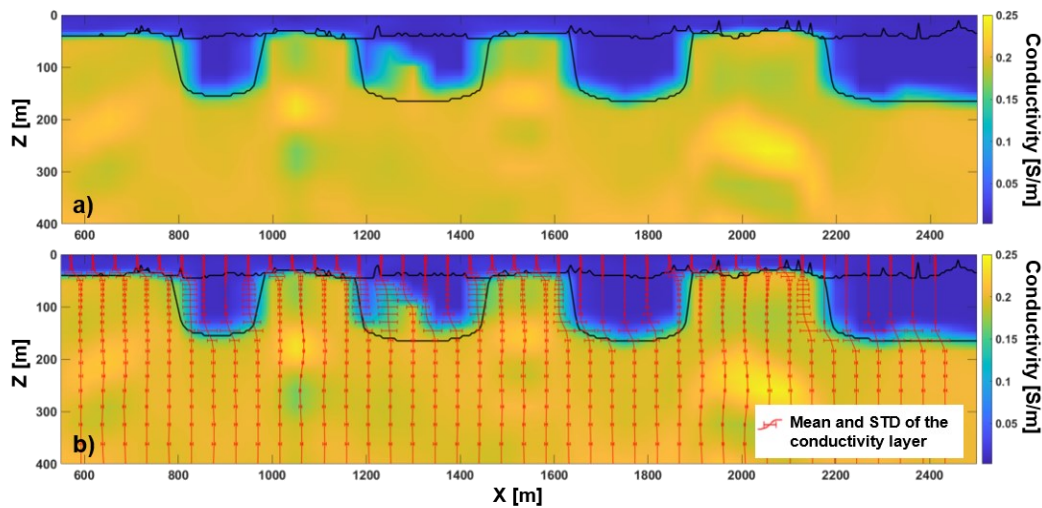

**Figure 13.** (**a**) Mean map of the 1D stochastic inversion of the 3D data (Figure 10c) associated to the conductivity model of Test 2 (Figure 10a shows one vertical section of that 3D conductivity model). In this case, realistic prior is used together with the modeling error assessment. The black lines show the locations of the interfaces to be reconstructed. (**b**) The same reconstructed conductivity distribution as in panel (**a**), but now with, superimposed, the mean and standard deviation vertical profiles.

## 4. Discussion

Are $10^5$ samples from the prior enough to guarantee the convergence of the stochastic inversions? What about the subset of 500 prior 3D realizations used for the assessment of the modeling error for the geological settings considered? Is the retrieved modeling error Gaussian as it should be to be able to use the proposed inversion scheme? In this section, we try to answer all these legitimate questions.

*4.1. About the Numerosity of the Prior Samples for the Convergency of the Stochastic Inversion*

Since the propositional scheme of the adopted Metropolis approach is based on a finite number of precalculated samples of the prior, it is important to establish if the abundance of those samples is sufficient to guarantee the convergence. In this respect, we run several inversions characterized by an increasing number of prior's samples. From Figures 14f–h, it is evident that, at least for the simple problem of Test 1, convergence is reached with a numerosity of the prior samples of a few thousand. Hence, an abundance of two orders of magnitude higher should reasonably be compatible with our goal. A similar conclusion can be deduced by considering the evolution of the correlation coefficient between the depth of the deepest interface inferred and the true one, as shown in Figure 15.

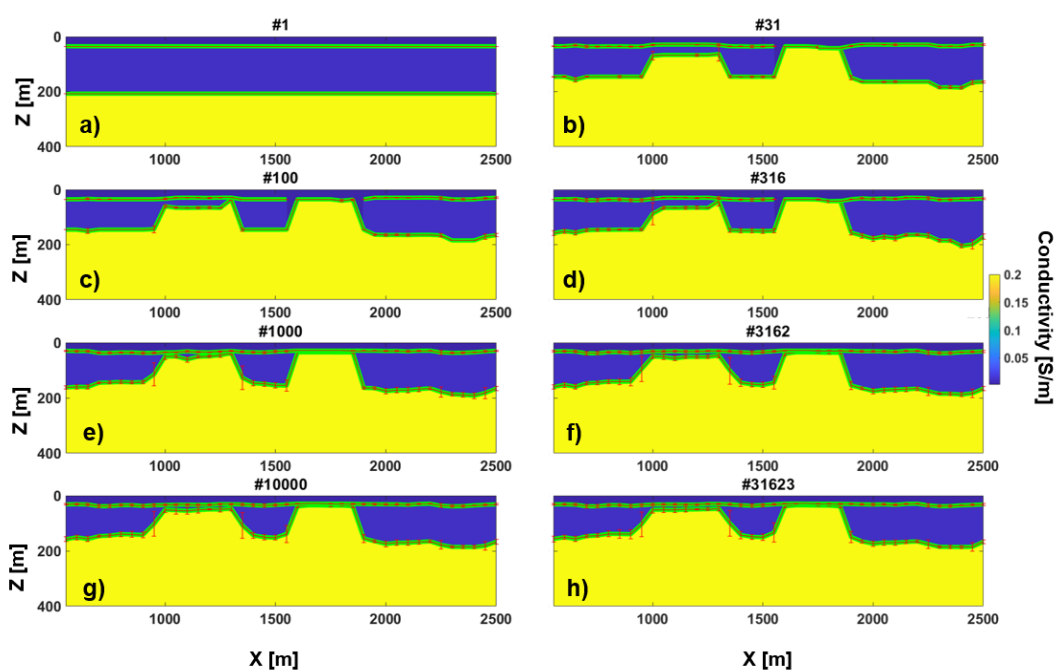

**Figure 14.** Result of the extended Metropolis inversion as a function of the numerosity of the considered precalculated prior samples – the title of each panel (**a** – **h**) reports that numerosity. Clearly, here, we are incorporating the information about the modeling error into the inversion of Test 1's data.

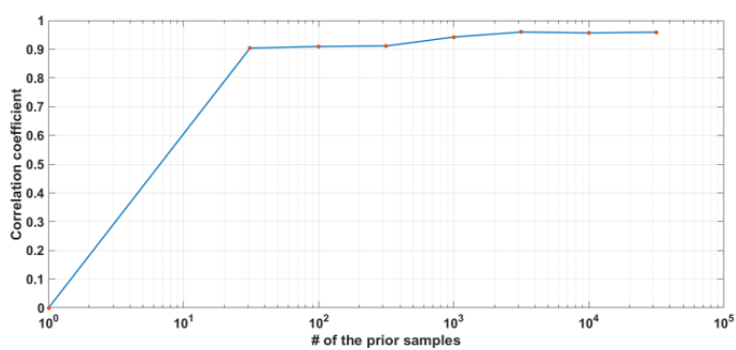

**Figure 15** Correlation coefficient between the retrieved and the true depth of the deepest layer as a function of the number of precalculated samples of the prior. Clearly, as for Figure 14, we are considering the stochastic inversion with modeling error assessment applied to Test 1's dataset.

*4.2. About the Numerosity of the Prior's Samples for the Estimation of the Modeling Error*

Still considering Test 1's dataset, we try to set up a strategy for the evaluation of the minimum number of realizations of the prior to be considered for an effective assessment of the modeling error. For all the tests performed in the present research, a maximum number of 500 models (similar to those in Figure 4) have been used to calculate the difference $D_{diff}$ and, in turn, the parameters defining the Gaussian probability density $\mathcal{N}(\mathbf{d}_\Delta, \mathbf{C}_\Delta)$ describing the modeling error. If we consider the behavior of the mean vector $\mathbf{d}_\Delta$ calculated for an increasing number of samples number $N_\Delta$, we can plot the results as in Figure 16a; from that figure, it appears evident that the assessment of $\mathbf{d}_\Delta$ reaches significant stability after considering a few hundred realizations of the prior.

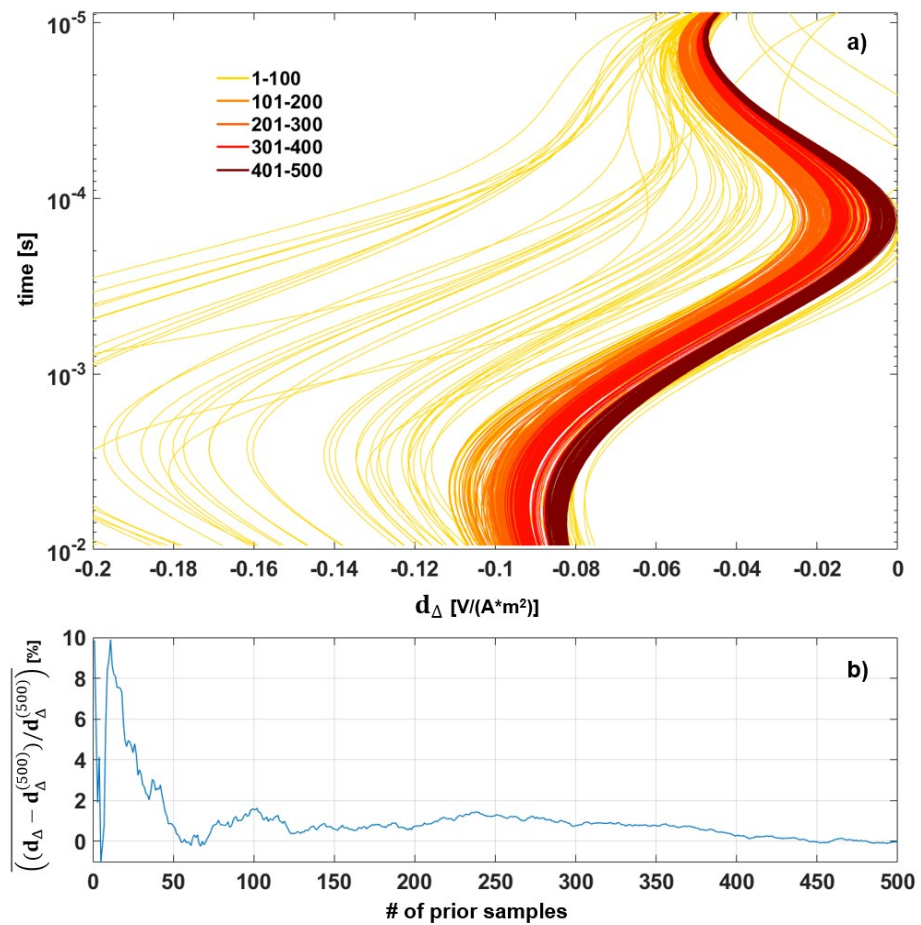

**Figure 16.** Mean of the modeling error as the number of the prior samples used for its estimation increases: (**a**) the mean $\mathbf{d}_\Delta$ vectors calculated for a number of prior samples ranging, for example, from 1 to 100 are represented by the solid yellow lines, whereas, considering another example, for a number of prior samples between 401 and 500, the same results for the $\mathbf{d}_\Delta$ vectors are plotted in dark red. (**b**) shows the behavior of the mean misfit between the estimated $\mathbf{d}_\Delta$ and our best assessment $\mathbf{d}_\Delta^{(500)}$ (based on 500 realizations).

A similar conclusion can be even more directly deduced by checking Figure 16b, in which the mean of the difference between $\mathbf{d}_\Delta$ and its best estimation $\mathbf{d}_\Delta^{(500)}$ (based on $N_\Delta = 500$ realizations) is plotted against the increasing numerosity $N_\Delta$: approximately after considering 400 samples, $\mathbf{d}_\Delta$ does not show significant variations.

Regarding the evolution of the covariance matrix $\mathbf{C}_\Delta$, we can, again, study how it changes over the number of prior samples. Figure 17 demonstrates, once more, that for the considered case, $N_\Delta =500$ should guarantee a reasonable estimation of the modeling error. In particular, Figure 17b shows that the estimation of $C_\Delta$ has already reached convergence after considering ~400 samples.

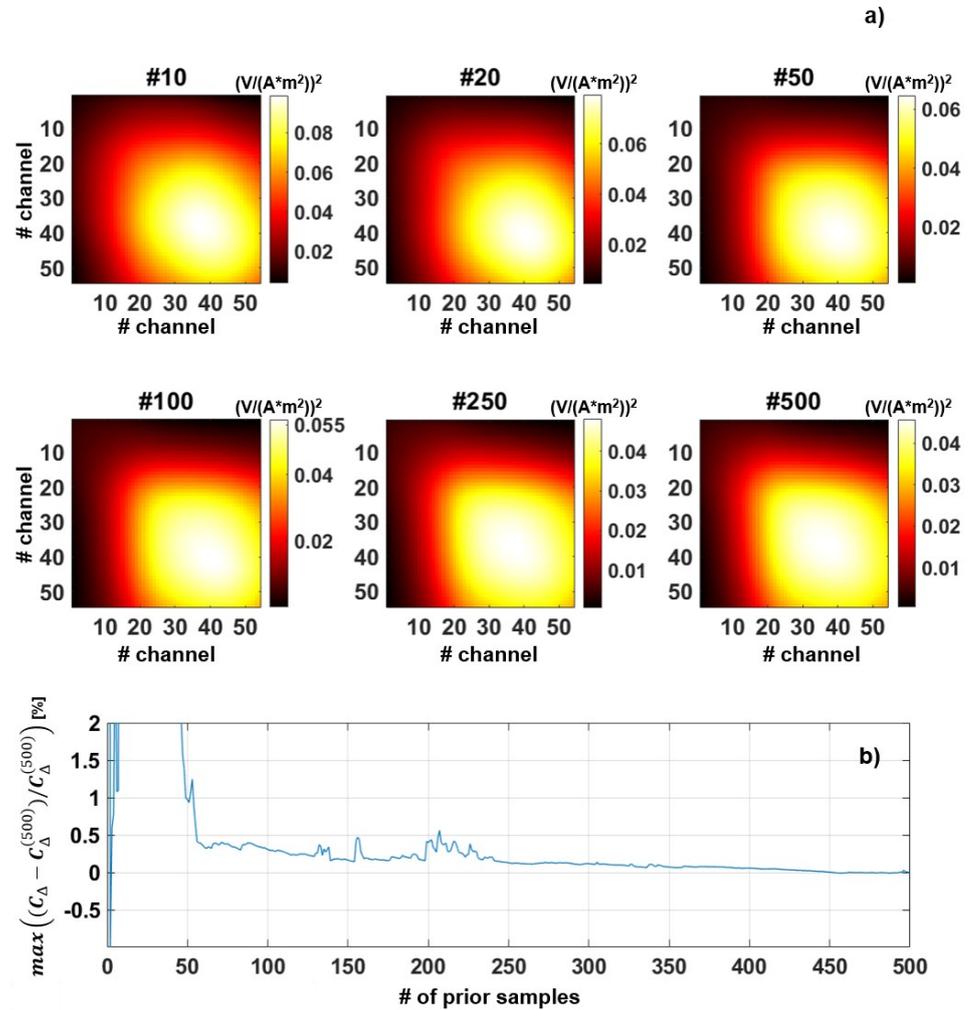

**Figure 17.** Covariance $C_\Delta$ of the modeling error as the number of prior samples used for its estimation increases: (**a**) different $C_\Delta$'s calculated for the number of prior samples indicated by the title of each subpanel (from $N_\Delta=10$ for the panel on the top-left to $N_\Delta=500$ for our best estimation on the bottom-right corner of panel (**a**)). (**b**) shows the behavior of the maximum misfit between all the components of the matrix $C_\Delta$ and the corresponding $C_\Delta^{(500)}$ based on our maximum number of realizations ($N_\Delta=500$).

*4.3. About the Gaussianity of the Modeling Error*

As mentioned before, the proposed approach is based on the assumption that the modeling error is actually Gaussian. In this section, we demonstrate that this working hypothesis is largely met. In this respect, each subpanel of Figure 18 shows one of the 54 histograms (one for each time-gate) of the corresponding component of the vector $D_{diff}$ of the difference between the 3D and 1D responses calculated for the 500 realizations used for the modeling error assessment. In particular, the red lines represent the Gaussian that is better fitting the experimental histograms, whereas the black lines show the Gaussian profile as inferred from the distribution $\mathcal{N}(\mathbf{d}_\Delta, C_\Delta)$. It is worth noticing the excellent agreement between the black and red curves, but, more than that, the fact that for the large majority of the time-gates the modeling error is indeed compatible with Gaussian distributions. Surprisingly, the histograms for the late channels (in particular those for $i$ going from ~49 to 54—i.e., basically the last row of Figure 18) show some sort of bimodal behavior whose possible justification is not evident. A possible reason might be connected with the finite size of the 3D simulation domain. However, this highly hypothetical guess will need to be investigated and verified.

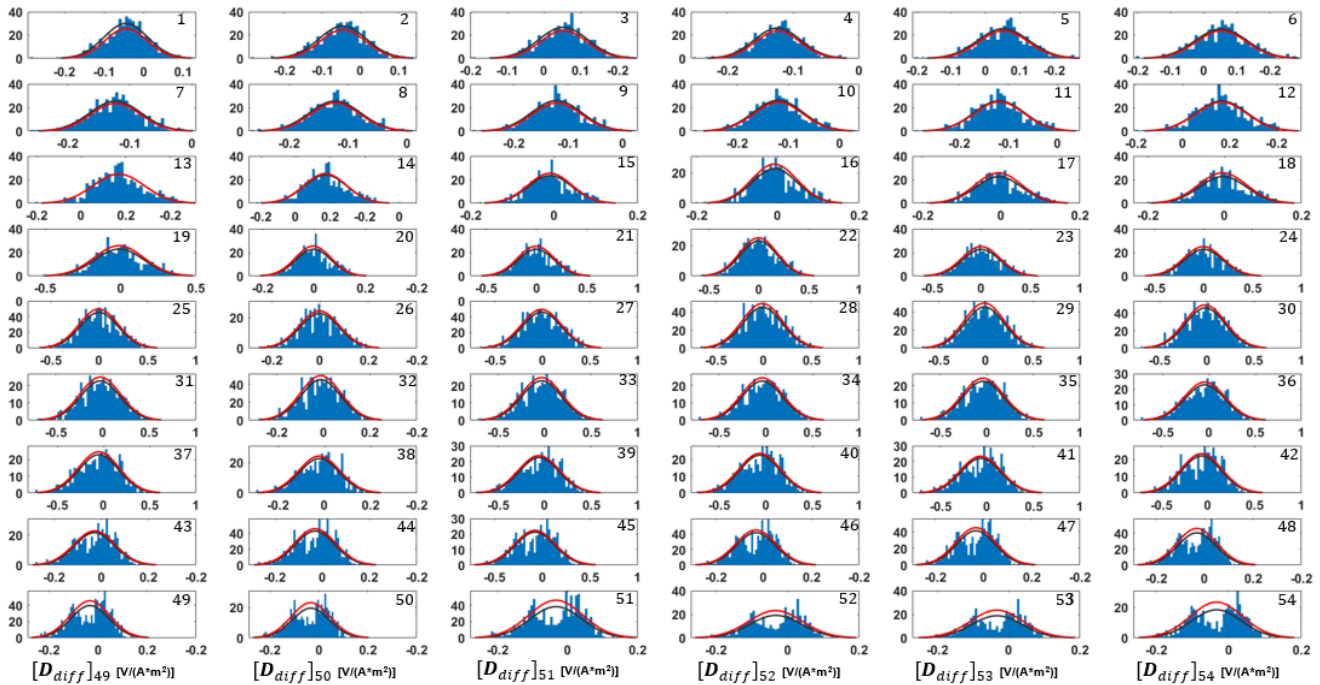

**Figure 18.** Each panel $j$ (with $j$, time-gate's index, varying from 1 to 54) shows the histogram of the corresponding component $[\mathbf{D}_{diff}]_j$ of the difference between the 3D and 1D responses for the 500 samples of the prior used for the modeling error estimation. The solid black line is the Gaussian curve better fitting the histogram, whereas the red line is the Gaussian derived by $\mathcal{N}(\mathbf{d}_\Delta, \mathbf{C}_\Delta)$.

## 5. Conclusions

Throughout this paper, two Tests are used to demonstrate the capabilities (and limitations) of stochastic inversion with geologically informed prior distribution to properly reconstruct the complex target conductivity distributions and the associated uncertainties.

In particular, we show how crucial the correct quantification of the modeling error (based on the prior choices) is. In our specific cases, we deal with ATEM data (mimicking the collection of VTEM measurements over geologies recalling glacial sedimentary environments typical, for example, of some regions in Northern Europe) inverted via a 1D stochastic approach with realistic prior and with the assessment of the corresponding modeling error. The conclusion is that, even in the case of stochastic approaches (already improving the results, for example, when compared with 1D deterministic approaches relying merely on the regularization term for including the prior knowledge about the targets), neglecting the fact that we are using a brute 1D approximation, instead of a more sophisticated 3D one, leads to unrealistically low levels of uncertainty in correspondence of those features that might turn out to be just artifacts. On the other hand, the assessment (and utilization) of the modeling error allows a more effective reconstruction of the true models and their associated reliability levels.

This research shows how the initial working hypotheses—concerning the minimum possible number of realizations defining the prior distribution and the numerosity of the smallest subset useful for an effective estimation of the modeling error—can be validated. Moreover, also the ansatz regarding the Gaussianity of the modeling error is largely verified.

Thus, the discussed workflow (presented and tested before on other kinds of data [47]) paves the way to the implementation of 1D probabilistic inversions of ATEM measurements capable to incorporate the complex pieces of geological information available and overcomes many of the difficulties connected with the utilization of efficient 1D approximations.

It is evident that the inclusion of the modeling error does not slow down the already available (and extremely fast) algorithms for 1D inversions (it does not matter if stochastic or deterministic). Hence, this approach will possibly remain useful also when fully 3D stochastic inversions will be practical; in fact, it will not be possible to consider any forward modeling tool perfect and, consequently, accounting for the modeling error will be, most likely, always beneficial.

It is also worth being highlighted that, at least for the investigated Tests, simply a few hundreds of 3D forward simulations are needed to retrieve a robust assessment of $\mathbf{d}_\Delta$ and $\mathbf{C}_\Delta$, and, actually, the same estimation for $\mathbf{d}_\Delta$ and $\mathbf{C}_\Delta$ can be used, in principle, in any survey characterized by similar conditions (i.e. similar prior). Hence, in those cases, the efforts for the calculation of $\mathbf{d}_\Delta$ and $\mathbf{C}_\Delta$ would impact merely the first survey and clearly would not increase with the size of the survey. On the contrary, a full 3D inversion requires at least a few tens of iterations (i.e. 3D calculations) for each sounding location; this easily results in thousands of expensive 3D forward simulations. From these considerations, it clearly appears how convenient the proposed approach is. On the other hand, it is probably true that severely 3D targets will be, in any case, poorly reconstructed by using 1D approaches; however, the proper inclusion of the modeling error will be always useful to correctly estimate the high uncertainty of the 1D reconstruction of 3D inclusions, whereas not taking into account the modeling error will, most likely, lead to wrong solutions that look (incorrectly) certain.

**Author Contributions:** Conceptualization, P.B., G.V. and T.M.H.; methodology, P.B., G.V. and T.M.H.; software, P.B., G.V. and T.M.H.; validation, P.B., G.V. and T.M.H.; formal analysis, P.B., G.V. and T.M.H.; investigation, P.B., G.V. and T.M.H.; resources, G.V.; data curation, P.B.; writing—original draft preparation, P.B. and G.V.; writing—review and editing, P.B., G.V. and T.M.H.; visualization, P.B. and G.V.; supervision, G.V. and T.M.H.; project administration, G.V. and T.M.H.; funding acquisition, G.V. and T.M.H. All authors have read and agreed to the published version of the manuscript

**Funding:** This research was partially funded: by the Independent Research Fund Denmark, grant number 7017-00160B; by the Italian Ministry of University and Research, through the initiative PON-RI 2014-2020, Asse I "Capitale Umano" – Azione I.1 "Dottorati innovativi con caratterizzazione industriale Ciclo XXXIII" – project: "GEOPROBARE: stochastic inversion of time-domain electromagnetic data" (CUP: F22J17000090007); by the University of Padova, through the research programme STARS@UNIPD – project: CHANGED – CHAracteriziNG pEatlands from Drones; by the Fondazione di Sardegna, through the initiative Progetti Biennali 2019 – project: "GEO-CUBE" (CUP: F72F20000250007).

**Acknowledgments:** Thanks are due to prof. Antonello Sanna, dr. Roberto Ricciu and dr. Giuseppe Desogus of the University of Cagliari (DICAAR) for providing most of the computer resources necessary to perform these tests.

**Conflicts of Interest:** The authors declare no conflict of interest.